\documentclass[%
reprint,
superscriptaddress,
 amsmath,amssymb,
 aps,
]{revtex4-1}

\usepackage{graphicx}
\usepackage{bm}
\usepackage[utf8]{inputenc}
\usepackage{amsmath}
\usepackage{float}
\usepackage{amsmath} 
\usepackage{booktabs}
\usepackage[x11names,table,xcdraw
]{xcolor}
\usepackage[colorinlistoftodos]{todonotes}

\usepackage{soul}

\begin{document}
\preprint{APS/123-QED}

\title{Excess velocity of magnetic domain walls close to the depinning field}

\author{Nirvana B. Caballero}
\affiliation{CNEA, CONICET, Centro Atómico Bariloche, Av. Bustillo 9500, 8400 S. C. de Bariloche, Río Negro, Argentina.}

\author{Iván Fernández Aguirre}
\affiliation{Instituto Balseiro, Univ. Nac. Cuyo - CNEA, Av. Bustillo 9500, 8400 S. C. de Bariloche, Rio Negro, Argentina.}

\author{Lucas J. Albornoz}
\affiliation{CNEA, CONICET, Centro Atómico Bariloche, Av. Bustillo 9500, 8400 S. C. de Bariloche, Río Negro, Argentina.}
\affiliation{Instituto Balseiro, Univ. Nac. Cuyo - CNEA, Av. Bustillo 9500, 8400 S. C. de Bariloche, Rio Negro, Argentina.}

\author{Alejandro B. Kolton}
\affiliation{CNEA, CONICET, Centro Atómico Bariloche, Av. Bustillo 9500, 8400 S. C. de Bariloche, Río Negro, Argentina.}
\affiliation{Instituto Balseiro, Univ. Nac. Cuyo - CNEA, Av. Bustillo 9500, 8400 S. C. de Bariloche, Rio Negro, Argentina.}

\author{Juan Carlos Rojas-S\'{a}nchez}
\affiliation{Unité Mixte de Physique, CNRS, Thales, Univ. Paris-Sud, Univ. Paris-Saclay, 91767 Palaiseau, France.}
\affiliation{Institut Jean Lamour, UMR CNRS 7198, Univ. de Lorraine, BP 70239, F-54506 Vandoeuvre, France.}

\author{Sophie Collin}
\affiliation{Unité Mixte de Physique, CNRS, Thales, Univ. Paris-Sud, Univ. Paris-Saclay, 91767 Palaiseau, France.}

\author{Jean Marie George}
\affiliation{Unité Mixte de Physique, CNRS, Thales, Univ. Paris-Sud, Univ. Paris-Saclay, 91767 Palaiseau, France.}

\author{Rebeca Diaz Pardo}
\affiliation{Laboratoire de Physique des Solides, CNRS, Univ. Paris-Sud, Universit\'e Paris-Saclay, 91405 Orsay, France.}

\author{Vincent Jeudy}
\affiliation{Laboratoire de Physique des Solides, CNRS, Univ. Paris-Sud, Universit\'e Paris-Saclay, 91405 Orsay, France.}

\author{Sebastian Bustingorry}
\affiliation{CNEA, CONICET, Centro Atómico Bariloche, Av. Bustillo 9500, 8400 S. C. de Bariloche, Río Negro, Argentina.}

\author{Javier Curiale}
\email{curiale@cab.cnea.gov.ar}
\affiliation{CNEA, CONICET, Centro Atómico Bariloche, Av. Bustillo 9500, 8400 S. C. de Bariloche, Río Negro, Argentina.}
\affiliation{Instituto Balseiro, Univ. Nac. Cuyo - CNEA, Av. Bustillo 9500, 8400 S. C. de Bariloche, Rio Negro, Argentina.}

\begin{abstract}

Magnetic field driven domain wall velocities in [Co/Ni] based multilayers thin films have been measured using polar magneto-optic Kerr effect microscopy. The low field results are shown to be consistent with the universal creep regime of domain wall motion, characterized by a stretched exponential growth of the velocity with the inverse of the applied field. Approaching the depinning field from below results in an unexpected excess velocity with respect to the creep law. We analyze these results using scaling theory to show that this speeding up of domain wall motion can be interpreted as due to the increase of the size of the deterministic relaxation close to the depinning transition. We propose a phenomenological model which allows to accurately fit the observed excess velocity and to obtain characteristic values for the depinning field $H_d$, the depinning temperature $T_d$, and the characteristic velocity scale $v_0$ for each sample.

\end{abstract}

\pacs{Valid PACS appear here}
\maketitle

\section{Introduction}

Nowadays, the fundamental relevance and potential technological impact of different magnetic textures, such as domain walls and skyrmions, are broadly acknowledged~\cite{RoadMap, Fert2017}. For example, ongoing research covers from skyrmions inspired applications~\cite{Nii2015, Huang2017, Yu2017} to domain walls response to in-plane magnetic fields in perpendicular magnetic anisotropy materials~\cite{Je2013, Hrabec2014, Lavrijsen2015, Vanatka2015, Belmeguenai2015, Jue2016natmat, Pham2016, Wells2017}. In particular, magnetic domain wall motion plays a key role on understanding magnetization reversal dynamics and on developing new magnetic based memory devices. This has strongly pushed research on domain wall dynamics during last years~\cite{Allwood2005, Hayashi2008, Parkin2008}.
It was soon realized that even a weak quenched disorder can have dramatic consequences in domain wall dynamics~\cite{nattermann_creep_law, ioffe_creep, lemerle_domainwall_creep}. Disorder competes with the domain wall elasticity, leading to a variety of non-trivial collective effects, including pinning, glassiness and dynamic phase transitions as a function of temperature and driving field~\cite{nattermann_creep_law, ioffe_creep, lemerle_domainwall_creep, chauve_creep_long, Ferre2013, Ferrero2013}. Despite this progress there are still many aspects deserving further attention, specially regarding the comparison between experiments and theoretical predictions for the different regimes. Among these, one of the most fundamental questions is to understand the measured mean stationary velocity of magnetic domain walls in response to an uniform driving field near the so called depinning threshold. In this work we directly address this basic issue.

\begin{figure*}[!ht]
\begin{center}
\def\svgwidth{2\columnwidth}
\begingroup%
  \makeatletter%
  \providecommand\color[2][]{%
    \errmessage{(Inkscape) Color is used for the text in Inkscape, but the package 'color.sty' is not loaded}%
    \renewcommand\color[2][]{}%
  }%
  \providecommand\transparent[1]{%
    \errmessage{(Inkscape) Transparency is used (non-zero) for the text in Inkscape, but the package 'transparent.sty' is not loaded}%
    \renewcommand\transparent[1]{}%
  }%
  \providecommand\rotatebox[2]{#2}%
  \ifx\svgwidth\undefined%
    \setlength{\unitlength}{1192bp}%
    \ifx\svgscale\undefined%
      \relax%
    \else%
      \setlength{\unitlength}{\unitlength * \real{\svgscale}}%
    \fi%
  \else%
    \setlength{\unitlength}{\svgwidth}%
  \fi%
  \global\let\svgwidth\undefined%
  \global\let\svgscale\undefined%
  \makeatother%
  \begin{picture}(1,0.3291559)%
    \put(0,0.01){\includegraphics[width=\unitlength]{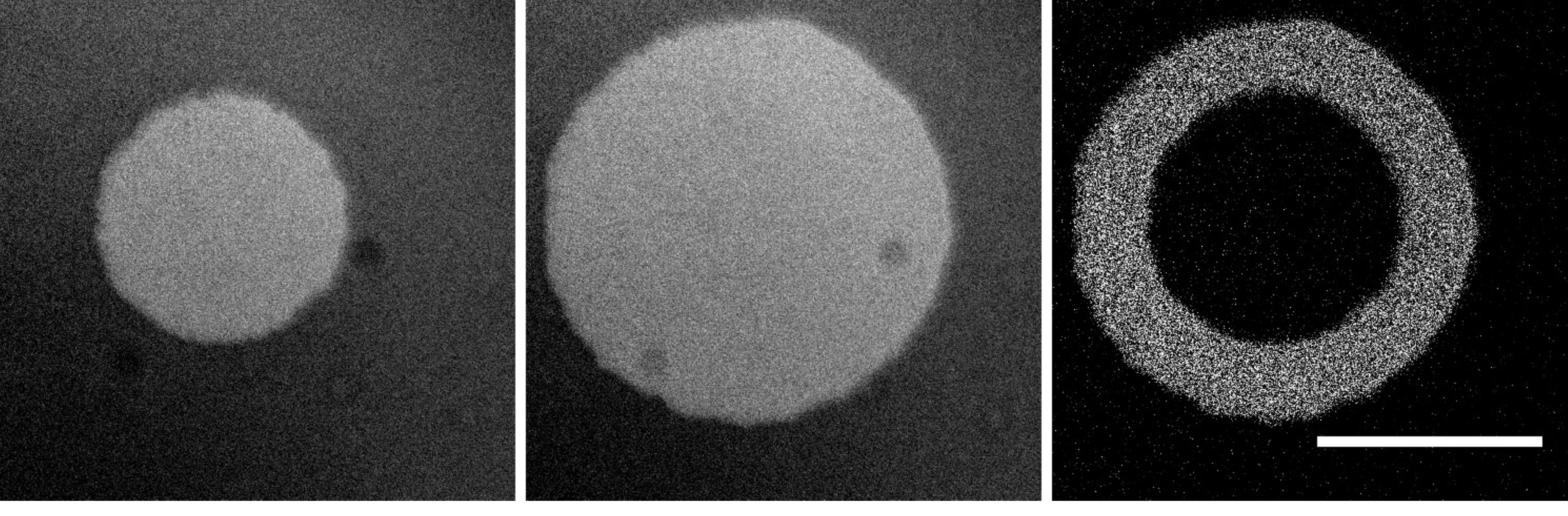}}%
    \put(0.89,0.045){\color[rgb]{1,0.33333333,0.33333333}\makebox(0,0)[lt]{\begin{minipage}{0.15349837\unitlength}\raggedright {\textcolor{white}{\large 50$\mu m$}}\end{minipage}}}%
    \put(0.02,0.32){\color[rgb]{0,0,0}\makebox(0,0)[lt]{\begin{minipage}{0.45591608\unitlength}\raggedright {\textcolor{white}{\large (a)}}\end{minipage}}}%
    \put(0.36,0.32){\color[rgb]{0,0,0}\makebox(0,0)[lt]{\begin{minipage}{0.26890616\unitlength}\raggedright {\textcolor{white}{\large (b)}}\end{minipage}}}%
    \put(0.69,0.32){\color[rgb]{0,0,0}\makebox(0,0)[lt]{\begin{minipage}{0.22277862\unitlength}\raggedright {\textcolor{white}{\large (c)}}\end{minipage}}}%
    \put(0.17,0.24){\color[rgb]{0,0,0}\makebox(0,0)[lt]{\begin{minipage}{0.36911866\unitlength}\raggedright \scalebox{1.5}{\textcolor{white}{\boldsymbol{$\odot$}}}\end{minipage}}}%
    \put(0.21,0.27){\color[rgb]{0,0,0}\makebox(0,0)[lt]{\begin{minipage}{0.64429885\unitlength}\raggedright \scalebox{1.5}{\textcolor{white}{\boldsymbol{$\otimes$}}}\end{minipage}}}%
    \put(0.625,0.275){\color[rgb]{0,0,0}\makebox(0,0)[lt]{\begin{minipage}{0.30201105\unitlength}\raggedright \scalebox{1.3}{\textcolor{white}{$\vec{H}_z$}}\end{minipage}}}%
    \put(0.6,0.27){\color[rgb]{0,0,0}\makebox(0,0)[lt]{\begin{minipage}{0.36911866\unitlength}\raggedright \scalebox{1.5}{\textcolor{white}{\boldsymbol{$\odot$}}}\end{minipage}}}%
    \put(0.235,0.275){\color[rgb]{0,0,0}\makebox(0,0)[lt]{\begin{minipage}{0.30201105\unitlength}\raggedright \scalebox{1.3}{\textcolor{white}{$\vec{M}$}}\end{minipage}}}%
  \end{picture}%
\endgroup%
\caption{PMOKE images for the CoNiTa sample. (a) Bubble domain after nucleation. Symbols $\boldsymbol{\odot}$ and $\boldsymbol{\otimes}$ indicate the direction of the magnetization in the two observed domains. (b) The same bubble domain after the first square field pulse of $30\,\mathrm{Oe}$ and $100\,\mathrm{ms}$. (c) Differential image corresponding to the two consecutive images (a) and (b).}
\label{fig:pmoke}
\end{center}
\end{figure*}
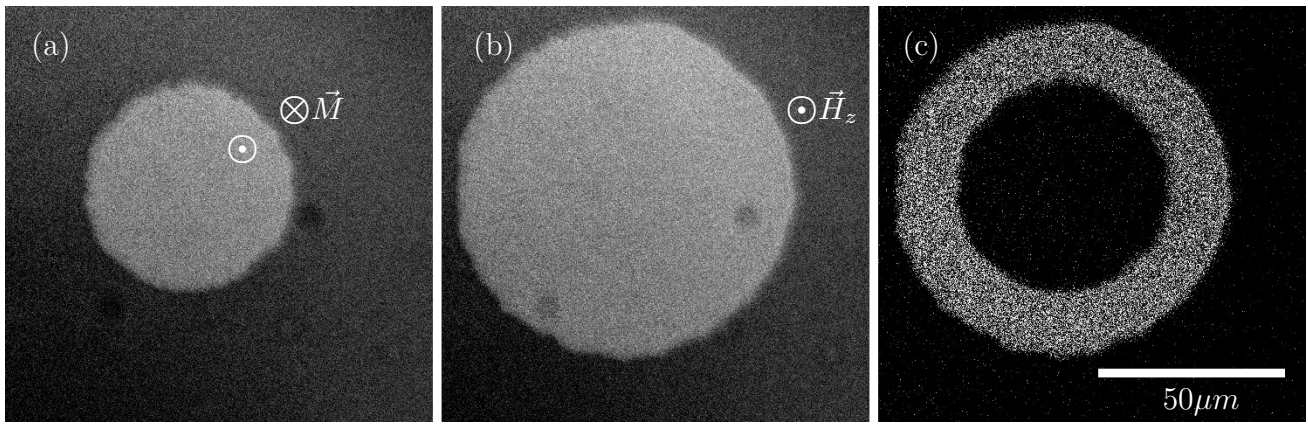

Typically, an external magnetic field is used to favor one particular domain orientation and can thus be taken as the exemplary driving force for domain wall motion. One of the key parameters is the characteristic depinning field $H_d$ separating slow from fast domain wall motion. Due to metastable states induced by quenched disorder, at zero temperature velocity is strictly zero below the depinning field $H_d$ and it is finite above it, defining the depinning transition. At large fields, $H \gg H_d$, velocity ultimately grows linearly with the field in the so-called flow regime. When decreasing the external field and close to the depinning field, $H \gtrsim H_d$, domain wall motion captures, through the temperature and field dependence of the velocity, essential universal signatures of the depinning transition at finite temperature~\cite{nowak_thermal_rounding, metaxas_depinning_thermal_rounding, bustingorry_thermal_rounding_epl, bustingorry_thermal_depinning_exponent, thermal_rounding_fitexp, Ferrero2013, Gorchon2014, DiazPardo2017}. Further reducing the external field below the depinning field, $H<H_d$, domain wall dynamics is controlled by thermal activation over effective energy barriers in the so-called creep regime~\cite{nattermann_creep_law, ioffe_creep, lemerle_domainwall_creep, chauve_creep_long, kolton_dep_zeroT_long, metaxas_depinning_thermal_rounding,Gorchon2014, Jeudy2016}. Within this regime, for any finite magnetic field, velocity is controlled by Arrhenius activation growing as $v = v_0e^{-\Delta E/k_BT}$, with $k_BT$ the thermal energy and a characteristic field-dependent disorder-induced effective energy barrier $\Delta E=(H/H_d)^{-\mu}-1$. The universal creep exponent $\mu$ depends on a few physical ingredients such as the dimension of the system, the extent of the elastic interaction (long- or short-range character) and the type of disorder correlations (random bond or random field, with short or long-range correlated pinning forces)~\cite{nattermann_creep_law, ioffe_creep, chauve_creep_long,kanda2010}.
For domain wall motion in ultra-thin ferromagnetic films $\mu=1/4$, corresponding to an elastic interface with short-range elasticity in a short-range correlated random-bond disorder.
The creep regime has been thoroughly studied during the past decades and still presents several challenges~\cite{Yamanouchi2007, Lee2011,lavrijsen2012, Moon2013, Lin2016, Jue2016natmat}.
An important unexplored issue is the physics of the prefactor $v_0$ in the velocity-force creep-law.
Usually, it is assumed to be constant while the forward displacement of domain walls from single non-overlapping events are predicted to be strongly field dependent~\cite{kolton_dep_zeroT_long} and to diverge close to the depinning threshold. A theoretical description containing such a force dependence for the prefactor, which is discussed in this paper, would have measurable experimental consequences and enrich the current understanding of slow DW dynamics.

Pt/Co/Pt ultrathin films have been used as the archetypal material system to study domain wall
dynamics~\cite{lemerle_domainwall_creep, metaxas_depinning_thermal_rounding, Emori2012, Moon2013, Gorchon2014}.
These systems permitted a deeper understanding of domain wall motion within the framework of disordered elastic systems, allowing us to differentiate material dependent parameters, such as the depinning field $H_d$ and the depinning temperature $T_d$, from universal characteristics as the critical exponents.
Due to its technological relevance materials offering other possibilities has been investigated. Among others, we mention the CoFeB family which present the lowest room temperature depinning field~\cite{Burrowes2013} and diluted magnetic semiconductors of the GaMnAs family showing low depinning fields but with a Curie temperature around 100~K \cite{Jeudy2016}.
Also interesting are ferrimagnetic systems such as TbFe and related materials presenting a compensation temperature close to room temperature~\cite{Pommier1994, Bang2014, Tono2015, Jeudy2016}.
Finally, [Co/Ni] multilayers were primarily studied as a promising system for future applications particularly due to its low propagation field and high spin-orbit coupling effect~\cite{Koyama2008, Tanigawa2009, Yamada2011, Koyama2011, Ryu2014, LeGall2015, Rojas2016}.  

In this work we present velocity-field characteristics for three different [Co/Ni] multilayers which show an unexpected deviation from the usual creep regime when approaching the depinning field $H_d$. We discuss in detail the observed behavior and propose a phenomenological model inspired by scaling theory and recent numerical simulations. Our model permits to fit relevant material dependent parameters corresponding to field, temperature and velocity scales.

\section{Domain wall velocity in [C\MakeLowercase{o}/N\MakeLowercase{i}] multilayers}
\label{sec:exp-results}

We present in this section velocity-field measurements in [Co/Ni] multilayers. Three samples were investigated: Pt(6)/[Co(0.2)/Ni(0.6)]$_{3}$/Ta(5), Pt(6)/[Co(0.2)/ Ni(0.6)]$_{3}$/Al(5), and Pt(6)/[Ni(0.6)/Co(0.2)]$_{3}$/Al(5). The numbers in parenthesis stand for thickness in nm. For a sake of simplicity, the samples are referred to as CoNiTa, CoNiAl, and NiCoAl, respectively in the following. The multilayers were grown on the same oxidized Si-SiO$_2$ substrates by DC magnetron sputtering and present a perpendicular magnetic anisotropy enhanced by the three periods of Co/Ni bilayer~\cite{Rojas2016}. 

Polar magneto-optical Kerr effect microscopy (PMOKE) has been used
to measure domain wall velocities. The most relevant features of the used microscope are the following: Olympus LMPLFLN series objectives (20x and 5x), high brightness red LED with a dominant wavelength 637nm, Glan-Thompson polarizers, a 14 bit CCD from \emph{QImaging Corp.}, and a K\"{o}hler configuration for the illumination. The external magnetic field was applied in the direction of the easy axis, i.e. perpendicular to the sample plane. 
Different coils and current sources have been used to generate magnetic field pulses of amplitudes up to $1$~kOe and duration spanning from 1${\mu}$s to several minutes.
To measure domain wall velocities the following protocol was used: first, sample magnetization is saturated in a given direction perpendicular to the sample. Afterwards, with a short and strong magnetic pulse one or more magnetic domains with magnetization pointing in the opposite direction, are nucleated. Finally, a series of square magnetic field pulses are applied and PMOKE images are taken between these pulses, thus capturing domain wall motion.
Domain wall velocity for a given magnetic field is then measured as the mean displacement of the domain wall, between successive images, over the pulse width. Typical PMOKE images of sample surface magnetization are shown in Fig.~\ref{fig:pmoke}. The magnetic contrast permits to directly observe in Fig.~\ref{fig:pmoke}(a) the domain structure just after nucleation process. Figure~\ref{fig:pmoke}(b) shows domain growth after a first magnetic field square pulse of amplitude 30~Oe and duration 100~ms is applied. From the differential image shown in Fig.~\ref{fig:pmoke}(c) the mean domain wall displacement is measured and the mean velocity is obtained.

\begin{figure}[ht!]
\begin{center}
\includegraphics[width=0.9\columnwidth]{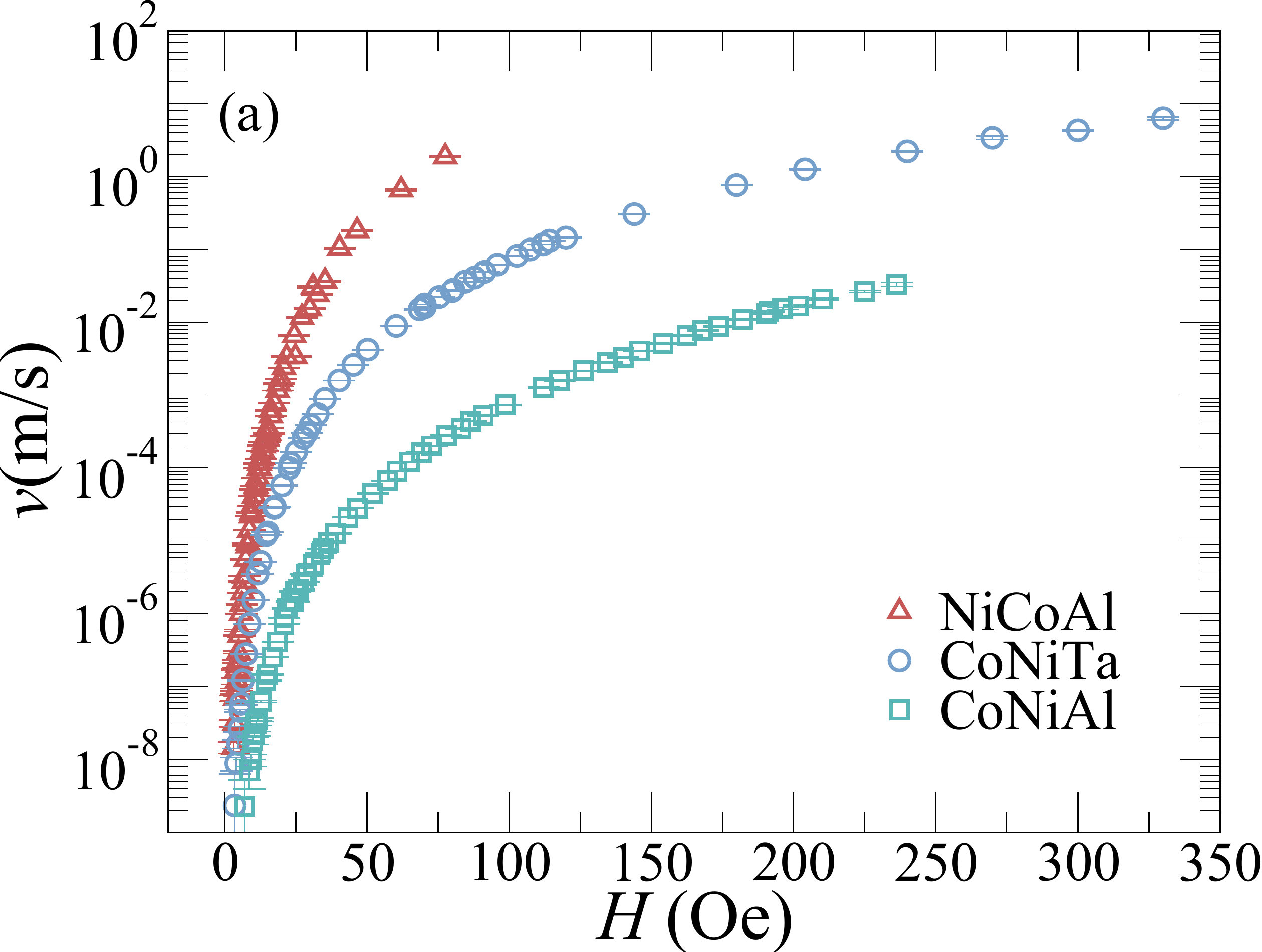}
\includegraphics[width=0.9\columnwidth]{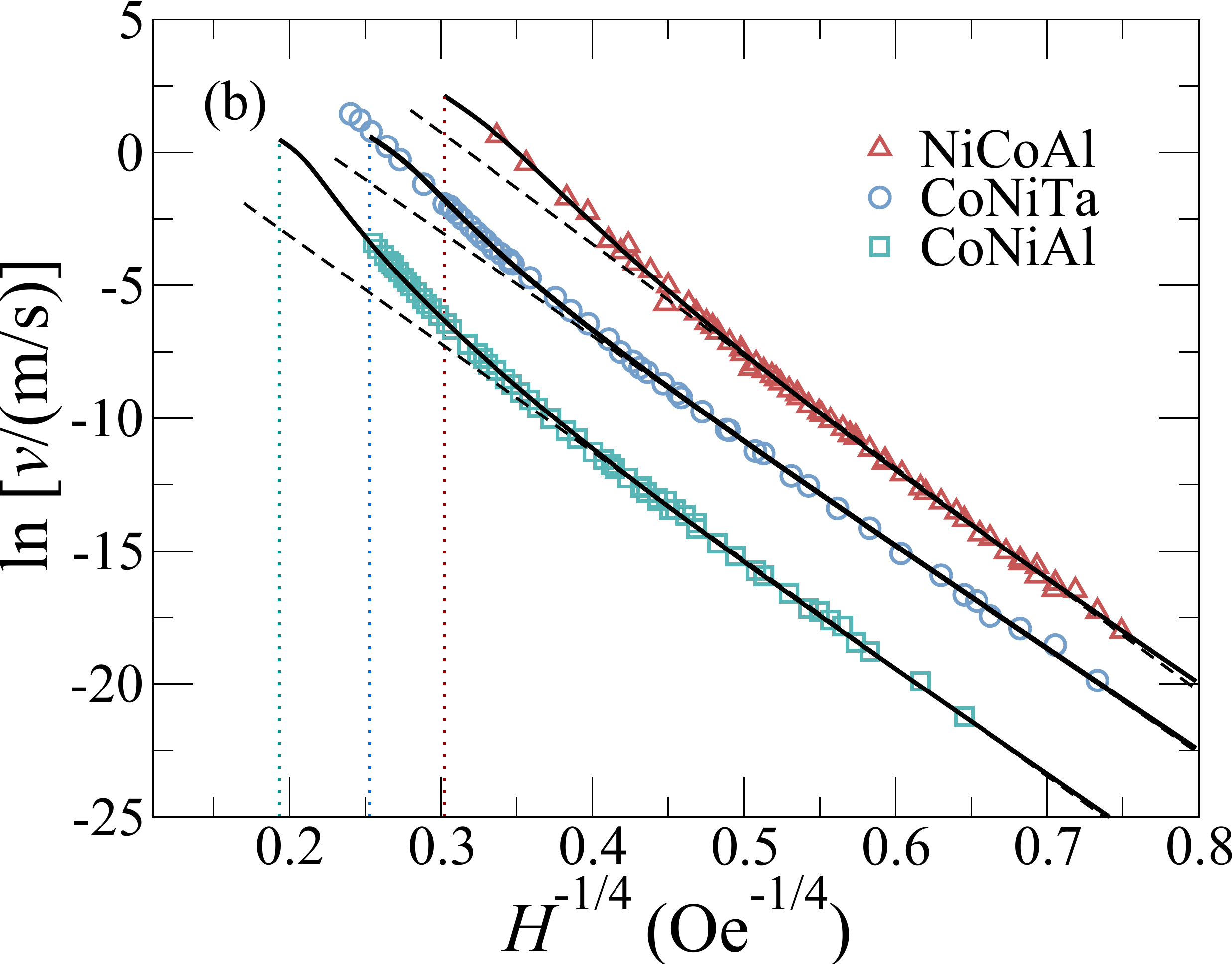}
\caption{Domain wall velocity as a function of magnetic field for [Co/Ni] multilayers: (a) Semilogarithmic scale and (b) creep plot, $\ln v$ against $H^{-1/4}$ (for simplicity, error bars were only included in (a)). Dashed lines in (b) are fits to the creep law $\ln v \sim H^{-1/4}$, which gives a good fit at low magnetic fields. The corrected creep law described in Sec.~\ref{sec:model} (continuous lines) gives a good description in an extended magnetic field range. Vertical dotted lines indicate the obtained depinning field $H_d$ for each curve. The corrected creep law was fit using fields $H<H_d$ (vertical dotted lines). }
\label{fig:M1M2M3}
\end{center}
\end{figure}

Velocity-field curves at room temperature are presented in Fig.~\ref{fig:M1M2M3}.
A semilogarithmic scale is used in Fig.~\ref{fig:M1M2M3}(a) to show that domain wall velocity covers 10 orders of magnitude for only two orders of magnitude of variation of the applied field. It can also be observed that the upper velocity limit is different for each sample. This is primarily related to the density of nucleation centers: the higher the density, the smaller the distance between domains and thus the maximum observable domain wall displacement is bounded by the coalescence of domains.
It is interesting to note that on samples with Al as capping layer (CoNiAl and NiCoAl), changing the order of the ferromagnetic bilayer changes the velocity almost 4 orders of magnitude for a given $H$.
Changing the capping layer also influences the velocity response as can be observed by comparing results for CoNiAl and CoNiTa.
Finally, our velocity measurements are similar to previous velocity-field results reported for Co/Ni based samples~\cite{LeGall2015}.
Taking the value $M_S = 300$~emu/cm$^3$ for the saturation magnetization, measured for our samples, and using $\alpha=0.05$ for the damping coefficient~\cite{Burrowes2013}, we have also estimated the Walker field $H_W=\alpha M_S/2 \approx 100$~Oe, which is smaller than the depinning field (see below).
However, we do not reach the velocity plateau observed at large fields in Refs.~\cite{Yamada2011, LeGall2015}.

In order to quantitatively study the domain wall dynamics on [Co/Ni] multilayers Fig.~\ref{fig:M1M2M3}(b) presents a creep plot, $\ln v$ against $H^{-1/4}$. For the lower magnetic fields, a linear behavior indicated by straight dashed lines presents very good agreement with the creep law. In contrast, for the higher magnetic field values, the velocity is found to be higher than expected from the creep law.

This excess velocity is encountered in other magnetic systems. It has been previously observed, but not highlighted, in a Pt/Co/Pt sample at room temperature in Ref.~\cite{Gorchon2014}, in Pt/Co/Pt magnetic wires of width 1.5$\mu m$ in Ref.~\cite{cayssol2004domain}, and also in epitaxially grown Co/Ni layer~\cite{LeGall2015}. The observed excess velocity contrasts with the phenomenological description for the finite temperature velocity-field response in disordered elastic systems~\cite{Ferrero2013}. In this scenario, when increasing the field, a simple crossover is expected between the standard creep law, where the velocity increases with the field faster than linear ($d^2v/dH^2>0$), and the above threshold fast flow regime, where the velocity reaches a linear behavior with $H$ ($d^2v/dH^2=0$). Therefore, the excess velocity corresponds to an increase of the velocity faster than the stretched exponential dependence with $H$ and then the phenomenological model is not enough to describe velocity values beyond the creep law. To explain the excess velocity effect we develop a model entirely based on the predictions of the creep theory. Corrections to the standard creep law derived from recent numerical studies of creep events~\cite{Ferrero2017} are essential to our model. As shown in the following sections, with this approach we are not only able to model quantitatively the excess velocity regime, but to predict its dependence with material parameters and the shape of the velocity-field characteristics, from low fields to fields near the depinning field, using a few fitting parameters.


\section{Excess velocity close to the depinning field}
\label{sec:model}

Inspired by previous numerical results~\cite{kolton_depinning_zerot2, kolton_dep_zeroT_long, Ferrero2017}, we develop here a phenomenological description of the excess velocity which is then compared to experimental results obtained for the [Co/Ni] multilayers and published previously for ultrathin Pt/Co/Pt films~\cite{Gorchon2014}.

\subsection{Phenomenological arguments and scaling theory}
\label{sec:corrected-model}

In the phenomenological creep model, it is assumed that the slow motion of domain walls at low fields and temperatures proceeds by nucleating optimal irreversible forward jumps over the effective energy barriers separating metastable states. These events, the so-called \textit{thermal nuclei}, are expected to be localized both in time and space: they occur in a very short time compared with the typical waiting time it takes to produce them in a given metastable state, and they have an optimal field-dependent size. In the first creep theories it was also implicitly assumed that these events were independent, and that metastable states were indistinguishable from equilibrium states.
However, simulations of creep motion and functional renormalization group calculations show that this picture is not correct. Indeed, metastable states are similar to equilibrium states only locally ~\cite{chauve_creep_long,kolton_depinning_zerot2, kolton_dep_zeroT_long} and activated events actually display complex spatio-temporal patterns and a power-law distribution of sizes~\cite{Ferrero2017}. Interestingly, the largest events (those that are in the cut-off of the size distribution) which dominate the velocity of the interface with their large waiting times, are almost independent and follow the same scaling laws early predicted for thermal nuclei~\cite{nattermann_creep_law,ioffe_creep}. This justifies the simple scaling approach we will follow for estimating the creep velocity from typical independent events.

\subsubsection{Creep-law}

The characteristic time to escape from a metastable configuration of the interface to another metastable state with lower energy via nucleation of the typical forward jump is given by the Arrhenius law
\begin{equation}
 \tau  = \tau_0 \, e^{\frac{{\Delta}E}{k_BT}},
 \label{eq:Tau}
 \end{equation}
with ${\Delta}E$ the typical effective energy barrier, $k_B$ the Boltzmann constant, $T$ the temperature, and $\tau_0$ a characteristic inverse attempt frequency. Using scaling~\cite{nattermann_creep_law, ioffe_creep} and functional renormalization group~\cite{chauve_creep_long} arguments it was shown that for small field values ($H \ll H_d$) the typical effective barrier scales as
\begin{equation}
\Delta E(H \to 0)=k_B T_d{\left(\frac{H}{H_d}\right)}^{-\mu},
\label{eq:DEold}
\end{equation}
with $k_B T_d$ a characteristic pinning energy scale depending on sample microscopic properties.
Remarkably, the predicted creep exponent $\mu$ is universal. For domain wall motion in ultra-thin ferromagnetic films $\mu = 1/4$, corresponding to the \textit{equilibrium universality class} of a one dimensional interface with short-ranged elasticity coupled to a short-range correlated random-bond disorder~\footnote{Fixing the equilibrium universality class do not fix the depinning universality class however. Anharmonic corrections to the elasticity for instance, do not change the equilibrium exponents but do change appreciably the depinning exponents~\cite{kolton_depinning_zerot2, kolton_dep_zeroT_long}.}.

Although Eq.~\eqref{eq:DEold} gives the dominant contribution to the effective barrier in the limit of small fields, corrections are generically expected at larger $H$. In particular, effective barriers should vanish when approaching the depinning field. Experimentally, it was shown~\cite{Gorchon2014, Jeudy2016} that the simple expression
\begin{equation}
\Delta E=k_B T_d\left[{\left(\frac{H}{H_d}\right)}^{-\mu}-1\right],
\label{eq:DE}
\end{equation}
describes well a large family of materials, from $H \ll H_d$ to $H\lesssim H_d$, for a considerable range of temperatures. It is worth noting that Eq.~\eqref{eq:DE} properly extrapolates to Eq.~\eqref{eq:DEold}, and shows that the effective barrier vanishes linearly when $H \to H_d$.

To compute the mean velocity $v$ of domain wall, we need to estimate the characteristic transverse displacement $u$ produced by individual creep events in addition to the characteristic times $\tau$ which is already given by Eqs.~\eqref{eq:Tau} and \eqref{eq:DE}. One usually considers that an event is triggered by the jump of a domain wall segment of length $L_{opt}$ (see Fig.~\ref{fig:lengths}) over an optimal energy barrier inducing a local displacement to a new metastable configuration. The length scale $L_{opt}$ acts then as the thermal nucleus triggering further forward deterministic motion of longitudinal length scale $L_{rel}$ towards the next metastable state, thus completing the whole creep event. The covered surface area of the complete event can be written $S_{eve}=L_{eve} u_{eve}$, where $u_{eve}$ is the transverse displacement and $L_{eve} \geq L_{opt}$ the longitudinal displaced length.
To get the center of mass displacement $u$, we assume a domain wall segment of length $L$ ($L > L_{eve}$) to undergo a number of events $n$ separated by a distance $\ell$ ($L\approx n\ell$) during the time scale $\tau$. The covered surface area is thus $uL=nL_{eve} u_{eve}$ which leads to:
\begin{equation}
u = \frac{S_{eve}}{\ell}.
\label{eq:uSeve}
\end{equation}

When the field is very small, $H \ll H_d$, the stretched exponential field-dependence of the time scale $\tau$, as shown below, dominates over less strong field-dependence of the domain wall displacement $u$. Therefore, as it is customary, a constant displacement value $u=u_0$ can be considered, resulting in the domain wall velocity given by $v=u_0/\tau$ and hence in the creep regime
\begin{equation}
v=v_0 e^{-\frac{T_d}{T}\left[ \left(\frac{H}{H_d}\right)^{-\mu}-1\right]},
\label{eq:FC}
\end{equation}
with $v_0=u_0/\tau_0$. This reduces to the well known creep-law
\begin{equation}
v=v_0 e^{-\frac{T_d}{T} \left(\frac{H}{H_d}\right)^{-\mu}},
\label{eq:FC2}
\end{equation}
when $H \to 0$.

\subsubsection{Field-dependent length scales}

Since the external magnetic field affects the interface energy landscape, it is natural to consider that the transverse displacement $u$, traveled by the interface when changing from one metastable state to another, has a dependence on the external magnetic field. We shall now consider the corrections to the velocity that might be originated on the field dependence of $u$.
To proceed, we first notice that, according to Eq.~\eqref{eq:uSeve}, the field-dependence of the size $S_{eve}$ of the magnetization reversal event should be specified. Assuming that domain walls can be described as self-affine objects \footnote{See Ref~\cite{Ferrero2013} for a review}, the transverse displacement $u$ is expected to present a power variation with the longitudinal size $L$: $u\sim L^{\zeta}$, with $\zeta$ the relevant roughness exponent, and then $S_{eve} \sim L_{eve}^{\zeta+1}$.

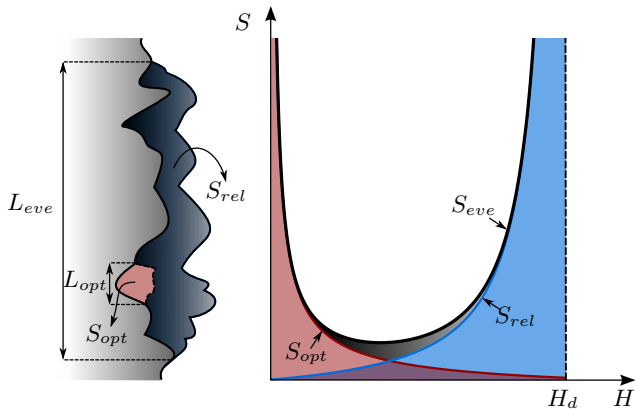
\begin{figure}[t!]
\begin{center}
\def\svgwidth{\columnwidth}
\begingroup%
  \makeatletter%
  \providecommand\color[2][]{%
    \errmessage{(Inkscape) Color is used for the text in Inkscape, but the package 'color.sty' is not loaded}%
    \renewcommand\color[2][]{}%
  }%
  \providecommand\transparent[1]{%
    \errmessage{(Inkscape) Transparency is used (non-zero) for the text in Inkscape, but the package 'transparent.sty' is not loaded}%
    \renewcommand\transparent[1]{}%
  }%
  \providecommand\rotatebox[2]{#2}%
  \ifx\svgwidth\undefined%
    \setlength{\unitlength}{793.91696167bp}%
    \ifx\svgscale\undefined%
      \relax%
    \else%
      \setlength{\unitlength}{\unitlength * \real{\svgscale}}%
    \fi%
  \else%
    \setlength{\unitlength}{\svgwidth}%
  \fi%
  \global\let\svgwidth\undefined%
  \global\let\svgscale\undefined%
  \makeatother%
  \begin{picture}(1,0.62920423)%
    \put(0.63899311,0.19554302){\color[rgb]{0,0,0}\makebox(0,0)[lb]{\smash{}}}%
    \put(0,0){\includegraphics[width=\unitlength,page=1]{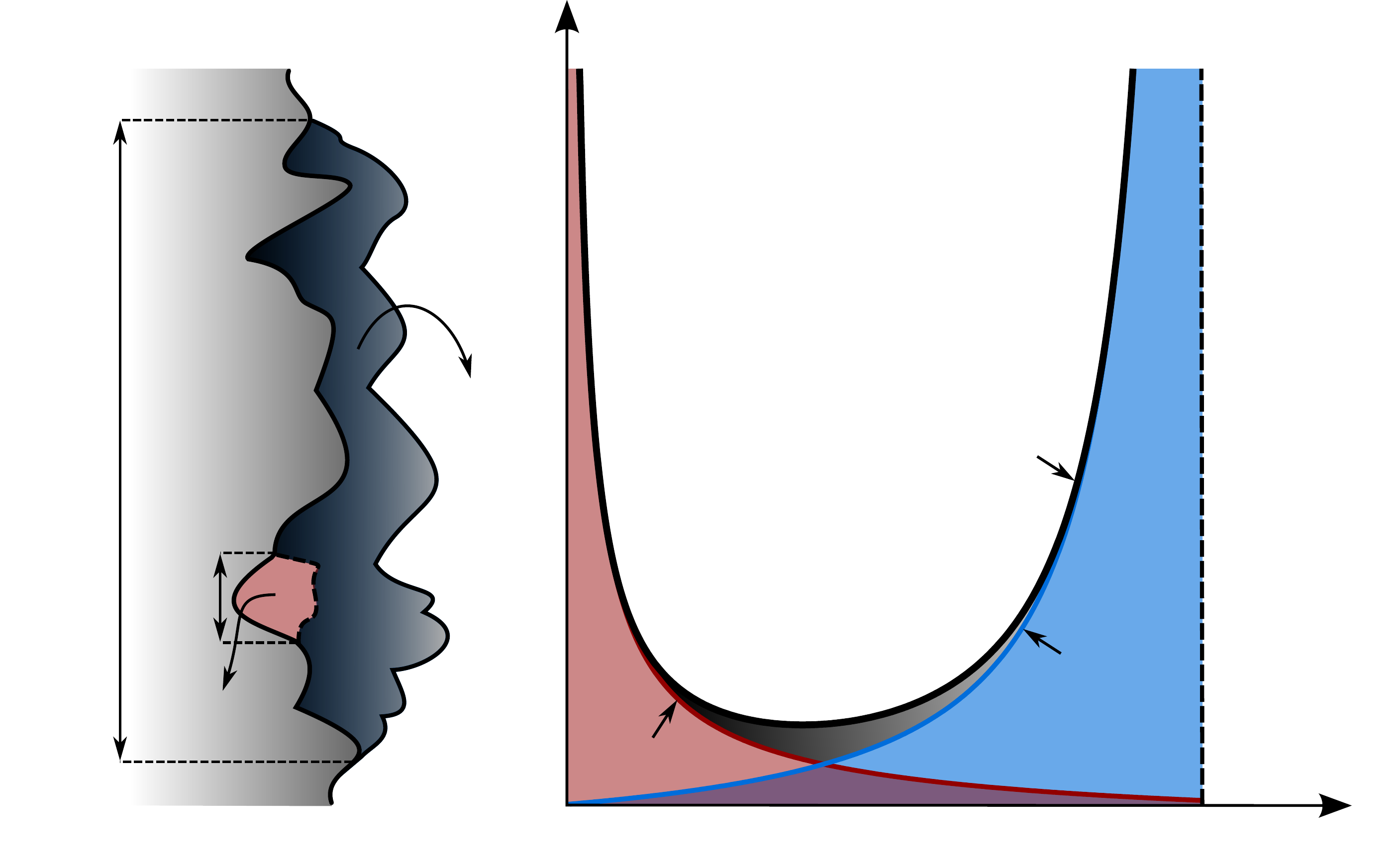}}%
    \put(0.83,0.00519576){\color[rgb]{0,0,0}\makebox(0,0)[lb]{\smash{\textit{$H_d$}}}}%
 
    \put(0.74983578,0.13267946){\color[rgb]{0,0,0}\makebox(0,0)[lb]{\smash{\textit{$S_{rel}$}}}}%

    \put(0.93121526,0.00519576){\color[rgb]{0,0,0}\makebox(0,0)[lb]{\smash{\textit{$H$}}}}%

    \put(0.42668179,0.07711477){\color[rgb]{0,0,0}\makebox(0,0)[lb]{\smash{\textit{$S_{opt}$}}}}%
    \put(0.68232245,0.30235522){\color[rgb]{0,0,0}\makebox(0,0)[lb]{\smash{\textit{$S_{eve}$}}}}%

    \put(0.12,0.13004505){\color[rgb]{0,0,0}\makebox(0,0)[lt]{\begin{minipage}{0.34724888\unitlength}\raggedright $S_{opt}$\end{minipage}}}%

    \put(0.305,0.35){\color[rgb]{0,0,0}\makebox(0,0)[lt]{\begin{minipage}{0.25695382\unitlength}\raggedright $S_{rel}$\end{minipage}}}%

    \put(0,0.30553357){\color[rgb]{0,0,0}\makebox(0,0)[lb]{\smash{\textit{$L_{eve}$}}}}%
    \put(0.085,0.18772872){\color[rgb]{0,0,0}\makebox(0,0)[lb]{\smash{\textit{$L_{opt}$}}}}%
    \put(0.35,0.60868452){\color[rgb]{0,0,0}\makebox(0,0)[lt]{\begin{minipage}{0.09068959\unitlength}\raggedright $S$\end{minipage}}}%
  \end{picture}%
\endgroup%
\caption{Characteristic length scales and corresponding sizes. $L_{eve}$ is the longitudinal length of an individual creep event. $L_{opt}$ corresponds to the optimal length associated to the effective energy barrier the domain wall should overcome in order to trigger a forward movement. $S_{opt}$ and $S_{rel}$ are the sizes of the corresponding magnetization reversed regions. On one hand, when $H \to 0$ one expects that $S_{opt} \sim L_{opt}^{\zeta_{eq}+1} \sim H^{-\nu_{eq}(\zeta_{eq}+1)}$, diverging at zero field. On the other hand, $S_{rel} \sim L_{rel}^{\zeta+1} \sim (H_d-H)^{-\nu(\zeta+1)}$ is diverging when approaching the depinning field from below. The size of the full event can thus be written as $S_{eve}=S_{opt}+S_{rel}$.}
\label{fig:lengths}
\end{center}
\end{figure}

In order to correct $u$ we should first recall the expected behavior for $S_{eve}$.
The relevant longitudinal length scales and areas are depicted in diagrams of Fig.~\ref{fig:lengths}. Scaling arguments~\cite{nattermann_creep_law, ioffe_creep, chauve_creep_long} and numerical simulations~\cite{Ferrero2017} show that at very small fields the relevant longitudinal length scale should be the one associated to the size of the optimal barrier the interface overcomes in order to produce a forward movement. This length scale is $L_{opt}$ and diverges with vanishing field as
\begin{equation}
\label{eq:Lopt}
 L_{opt} = L_c \left(\frac{H}{H_d} \right)^{-\nu_{eq}},
\end{equation}
with $L_c$ the Larkin length and $\nu_{eq} = 1/(2-\zeta_{eq})=3/4$ the correlation length equilibrium exponent, where $\zeta_{eq}=2/3$ is the equilibrium roughness exponent corresponding to the equilibrium universality class of a one-dimensional interface in a short-range correlated potential produced by random-bond disorder~\cite{kardar_review_lines}; these exponents yield $\mu = \nu_{eq} (2\zeta_{eq}-1)=1/4$~\cite{nattermann_creep_law, ioffe_creep}.

Remarkably, this power-law divergence of $L_{opt}$ is supported by experimental evidence of a dimensional crossover of DW dynamics in nanowires~\cite{Kim2009}.
The transverse displacement associated with $L_{opt}$, can be written as
\begin{equation}
u_{opt} = r_f (L_{opt}/L_c)^{\zeta_{eq}},   
\end{equation}
with $r_f$ the correlation length of the pinning force acting on the domain wall~\cite{chauve_creep_long}. The area of this contribution is thus given by 
\begin{equation}
\label{eq:Sopt}
S_{opt}=u_{opt} L_{opt} = L_c r_f  (L_{opt}/L_c)^{\zeta_{eq}+1}.
\end{equation}

When approaching the depinning field from below, deterministic relaxation events, similar to depinning avalanches, can be triggered by each thermal nucleus giving an additional contribution $L_{rel}$. Numerical simulations show that this length diverges when approaching the depinning field from below as $L_{rel} \sim (H_d-H)^{-\nu}$, with $\nu$ the depinning correlation length exponent~\cite{kolton_depinning_zerot2, kolton_dep_zeroT_long, Ferrero2017}. In terms of the reduced field $\delta h \equiv 1-H/H_d$ we can write
\begin{equation}
\label{eq:Lreldiv}
 L_{rel} \approx  L_c \delta h^{-\nu},
\end{equation}
for the longitudinal size of the relaxation event, with $\nu=1/(2-\zeta)$ and $\zeta$~\cite{Ferrero_PRE_2013_depinning} the depinning correlation length and depinning roughness exponents. This power law variation represents the relevant contribution close to $H_d$ and should be corrected when approaching the $H \to 0$ limit. In order to give a negligible contribution in the small field limit, where $L_{opt}$ is the dominant scale, we propose to recast the behavior of $L_{rel}$ in the form
\begin{equation}
\label{eq:Lreldiv1}
 L_{rel}=L_c \left( \delta h^{-\nu} -1 \right),
\end{equation}
that includes the limit $L_{rel} \to 0$ when $H \to 0$.
In terms of the relaxation length, the transverse displacement and the area of the relaxation contribution become
\begin{equation}
u_{rel} \approx r_f(L_{rel}/L_c)^{\zeta}  
\end{equation}
and
\begin{equation}
\label{eq:SrelLrel}
S_{rel} \approx L_c r_f(L_{rel}/L_c)^{\zeta+1},
\end{equation}
respectively.

\subsubsection{Corrected phenomenological model}

We are now in position to estimate the velocity prefactor $u/\tau_0 \sim S_{eve}$, using that $S_{eve}=S_{rel}+S_{opt}$. The first step towards the formulation of a corrected phenomenological velocity model is to realize that at very small fields, $H \ll H_d$, we have $S_{eve} \approx S_{opt} = r_f L_c (H/H_d)^{-\nu_{eq}(\zeta_{eq}+1)}$, according to Eqs.~\eqref{eq:Lopt} and \eqref{eq:Sopt}.
However, at very small fields the stretched exponential field dependence of the time scale dominates over this power-law correction for the prefactor. Therefore, in the very small field regime the size of the event can be conveniently approximated by a constant value, $S_{eve} \approx S_{opt} \approx S_0$. This yields for a constant velocity prefactor $v_0 = (S_0/\ell)/\tau_0$ and the creep-law Eq.~\eqref{eq:FC2} is recovered.

When increasing the field, the exponential term dominates over the $S_{opt}$ field dependence, which is decreasing with increasing field, and we can then safely consider $S_{opt} \approx S_0$ up to $H_d$. However, the relaxation contribution $S_{rel}$ increases with increasing field (see Fig. \ref{fig:lengths}) and starts competing with the stretched exponential field dependence from the time scale. One should then consider the relaxation correction to the velocity for fields larger than a given field $H_r$, to be defined below. The excess contribution to the full event as originated from the relaxation contribution can be considered by writing the size of the event as
\begin{equation}
\label{eq:velocity-prefactor}
 S_{eve} = S_0 + S_{rel} = S_0 \left( 1+\frac{S_{rel}}{S_0} \right),
\end{equation}
with $S_{rel} \to 0$ when $H \to 0$. Using Eqs.~\eqref{eq:Lreldiv1} and \eqref{eq:SrelLrel} $S_{eve}$ can be expressed as a function of the reduced field $\delta h$,
\begin{equation}
\label{eq:Sevedh}
 S_{eve} = S_0 \left[ 1+ p \left(\delta h^{-\nu}-1 \right)^{\zeta+1} \right],
\end{equation}
with $p=L_c r_f/S_0$.
Now, using Eq.~\eqref{eq:uSeve} to write the transverse displacement $u$ in terms of $S_{eve}$ results in the velocity given by
\begin{equation}
\label{eq:veldh}
 v=v_0 \left[ 1 + p \left( \delta h^{-\nu}-1 \right)^{\zeta+1} \right] e^{-\frac{T_d}{T}\left[ \left(1-\delta h\right)^{-\mu}-1\right]},
\end{equation}
where we have written $H/H_d=1-\delta h$ in the exponential. Equation~\eqref{eq:veldh} models the excess velocity observed when approaching the depinning field. A rough estimation for the field $H_r$ can be obtained by approximating the velocity prefactor in Eq.~\eqref{eq:veldh} as $v_0 p \delta h^{-\nu(\zeta+1)}$. This approximation for the prefactor can be recast in the $\delta h$-dependence of the exponential as $\exp\{-[(1-\delta h)^{-\mu}-1]/t-\nu(\zeta+1) \ln \delta h \}$, with $t=T/T_d$ a reduced temperature.
Then $H_r$ is obtained when the two terms in the exponential are of the same order:
\begin{equation}
 -\frac{1}{t} \left[ \left( 1-\delta h_r \right)^{-\mu}-1\right] \approx \nu(\zeta+1) \ln \delta h_r,
\end{equation}
giving an implicitly approximate expression for $H_r=H_d(1-\delta h_r)$.

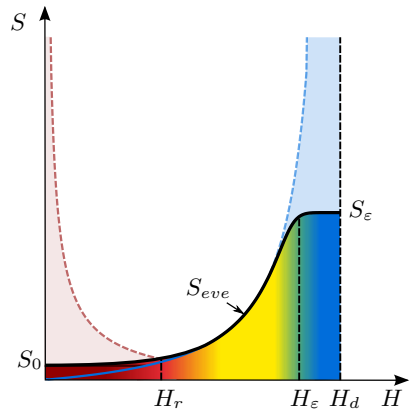
\begin{figure}[th!]
\begin{center}
\def\svgwidth{\columnwidth}
\begingroup%
  \makeatletter%
  \providecommand\color[2][]{%
    \errmessage{(Inkscape) Color is used for the text in Inkscape, but the package 'color.sty' is not loaded}%
    \renewcommand\color[2][]{}%
  }%
  \providecommand\transparent[1]{%
    \errmessage{(Inkscape) Transparency is used (non-zero) for the text in Inkscape, but the package 'transparent.sty' is not loaded}%
    \renewcommand\transparent[1]{}%
  }%
  \providecommand\rotatebox[2]{#2}%
  \ifx\svgwidth\undefined%
    \setlength{\unitlength}{577.3984359bp}%
    \ifx\svgscale\undefined%
      \relax%
    \else%
      \setlength{\unitlength}{\unitlength * \real{\svgscale}}%
    \fi%
  \else%
    \setlength{\unitlength}{\svgwidth}%
  \fi%
  \global\let\svgwidth\undefined%
  \global\let\svgscale\undefined%
  \makeatother%
  \begin{picture}(1,0.62920423)%
    \put(0,0){\includegraphics[width=\unitlength]{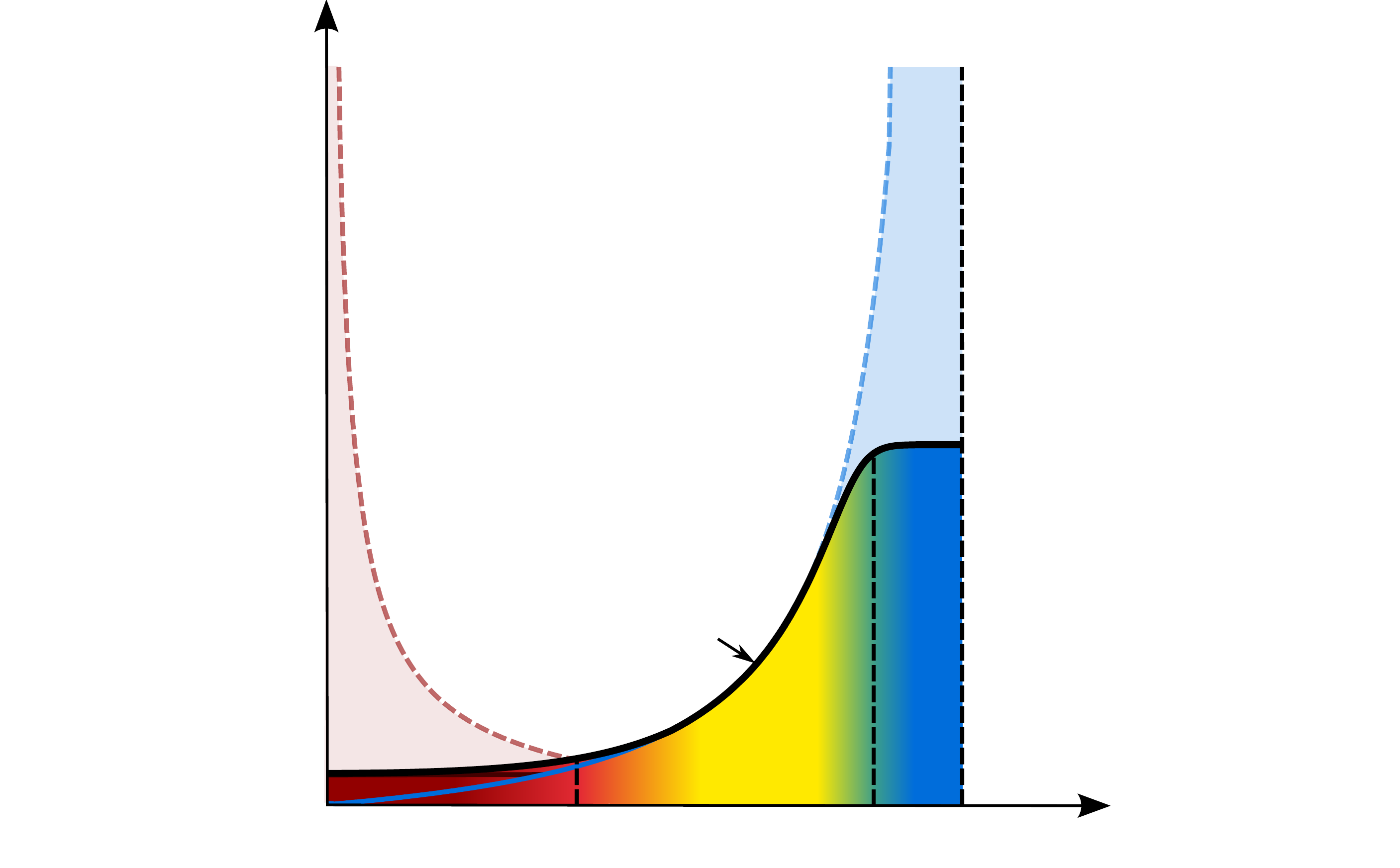}}%
    \put(0.18,0.60868452){\color[rgb]{0,0,0}\makebox(0,0)[lt]{\begin{minipage}{0.09068959\unitlength}\raggedright $S$\end{minipage}}}%
    \put(0.19,0.09){\color[rgb]{0,0,0}\makebox(0,0)[lt]{\begin{minipage}{0.09068959\unitlength}\raggedright $S_0$\end{minipage}}}%
    \put(0.45,0.2){\color[rgb]{0,0,0}\makebox(0,0)[lt]{\begin{minipage}{0.25007038\unitlength}\raggedright $S_{eve}$\end{minipage}}}%
    \put(0.7,0.32){\color[rgb]{0,0,0}\makebox(0,0)[lt]{\begin{minipage}{0.37037451\unitlength}\raggedright $S_{\varepsilon}$\end{minipage}}}%
        \put(0.40,0.00519576){\color[rgb]{0,0,0}\makebox(0,0)[lb]{\smash{\textit{$H_r$}}}}%
        \put(0.61,0.00519576){\color[rgb]{0,0,0}\makebox(0,0)[lb]{\smash{\textit{$H_{\varepsilon}$}}}}%
        \put(0.67,0.00519576){\color[rgb]{0,0,0}\makebox(0,0)[lb]{\smash{\textit{$H_d$}}}}%
        \put(0.75,0.00519576){\color[rgb]{0,0,0}\makebox(0,0)[lb]{\smash{\textit{$H$}}}}%
  \end{picture}%
\endgroup%
\caption{Proposed characteristic dependence of the relaxation size $S_{eve}$ with magnetic field. $S_{eve}$ has two contributions: $S_{opt}$ and $S_{rel}$ (see Fig.~\ref{fig:lengths}). $S_{eve}$ goes to a constant value $S_0$ at low fields. This is a good approximation since at low fields ($H<H_r$) the domain wall velocity is dominated by the exponential time scale and at higher fields ($H>H_r$) it is dominated by $S_{rel}$. A saturation of $S_{rel}$ is considered in order to arrest the divergence of the domain wall velocity near the depinning field. For $H < H_\varepsilon$, $S_{rel}$ is barely changed, but for $H>H_\varepsilon$ a saturation to $S_\varepsilon$ is reached.}
\label{fig:saturation}
\end{center}
\end{figure}

However, although we stated that $S_{rel}$ should diverge close to $H_d$, this would imply a divergence of $u$ and then of the velocity when approaching $H_d$ from below. This divergence should be arrested by a proper cut-off coming from experimental relevant length scales (system size, average distance between strong pinning centers, typical distance between relaxation events, etc.). Naming $L_\varepsilon$ to this longitudinal cut-off length, the corresponding size is
\begin{equation}
S_\varepsilon = L_c r_f (L_\varepsilon/L_c)^{\zeta+1}.
\end{equation}
Using Eq.~\eqref{eq:Lreldiv1}, the value $L_\varepsilon$ is reached when the reduced field is $\delta h_\varepsilon=\varepsilon=(L_\varepsilon/L_c+1)^{-1/\nu}$, corresponding to a field value $H_\varepsilon=H_d(1-\varepsilon)$.

To model the divergence of $S_{rel}$ and its corresponding cut-off we propose to replace $\delta h^{-\nu} \to (\delta h^m + \varepsilon^m)^{-\nu/m}$, which amounts to use
\begin{equation}
\label{eq:Seve}
 \frac{S_{eve}}{S_0} = 1 + p \left[ \left(\delta h^m + \varepsilon^m \right)^{-\nu/m}-1 \right]^{\zeta+1},
\end{equation}
instead of Eq.~\eqref{eq:Sevedh}. This model represents a crossover function where the width of the crossover is controlled by the parameter $m$ and is such that
\begin{equation}
\label{eq:Srellim}
 S_{rel} \sim \left\{
\begin{array}{ccc}
 (\delta h^{-\nu}-1)^{\zeta+1} & \mathrm{for} & \varepsilon \ll \delta h, \\
 (\varepsilon^{-\nu}-1)^{\zeta+1} & \mathrm{for} & \delta h \ll \varepsilon,
\end{array}
\right.
\end{equation}
independently of $m$. We fixed the value $m=4$ in our model, but using larger values of $m$ does not significantly change the velocity behavior.

Equation~\eqref{eq:Seve} reflects the fact that $S_{rel} \to 0$ when $H \to 0$. When the field is increased, $S_{rel}$ starts diverging while approaching $H_d$, but when the field is too close to $H_d$ ($\delta h$ smaller than $\varepsilon$ or, equivalently, $H$ larger than $H_\varepsilon$) it saturates to $S_\varepsilon$. The proposed behavior for $S_{eve}$ is outlined in Fig.\ref{fig:saturation}.

\begin{figure}[th!]
\begin{center}
\def\svgwidth{0.8\columnwidth}
\begingroup%
  \makeatletter%
  \providecommand\color[2][]{%
    \errmessage{(Inkscape) Color is used for the text in Inkscape, but the package 'color.sty' is not loaded}%
    \renewcommand\color[2][]{}%
  }%
  \providecommand\transparent[1]{%
    \errmessage{(Inkscape) Transparency is used (non-zero) for the text in Inkscape, but the package 'transparent.sty' is not loaded}%
    \renewcommand\transparent[1]{}%
  }%
  \providecommand\rotatebox[2]{#2}%
  \ifx\svgwidth\undefined%
   \setlength{\unitlength}{793.91696167bp}%
    \ifx\svgscale\undefined%
      \relax%
    \else%
      \setlength{\unitlength}{\unitlength * \real{\svgscale}}%
    \fi%
  \else%
    \setlength{\unitlength}{\svgwidth}%
  \fi%
  \global\let\svgwidth\undefined%
  \global\let\svgscale\undefined%
  \makeatother%
  \begin{picture}(1,0.62920423)%
  \put(0,0){\includegraphics[width=\unitlength,page=1]{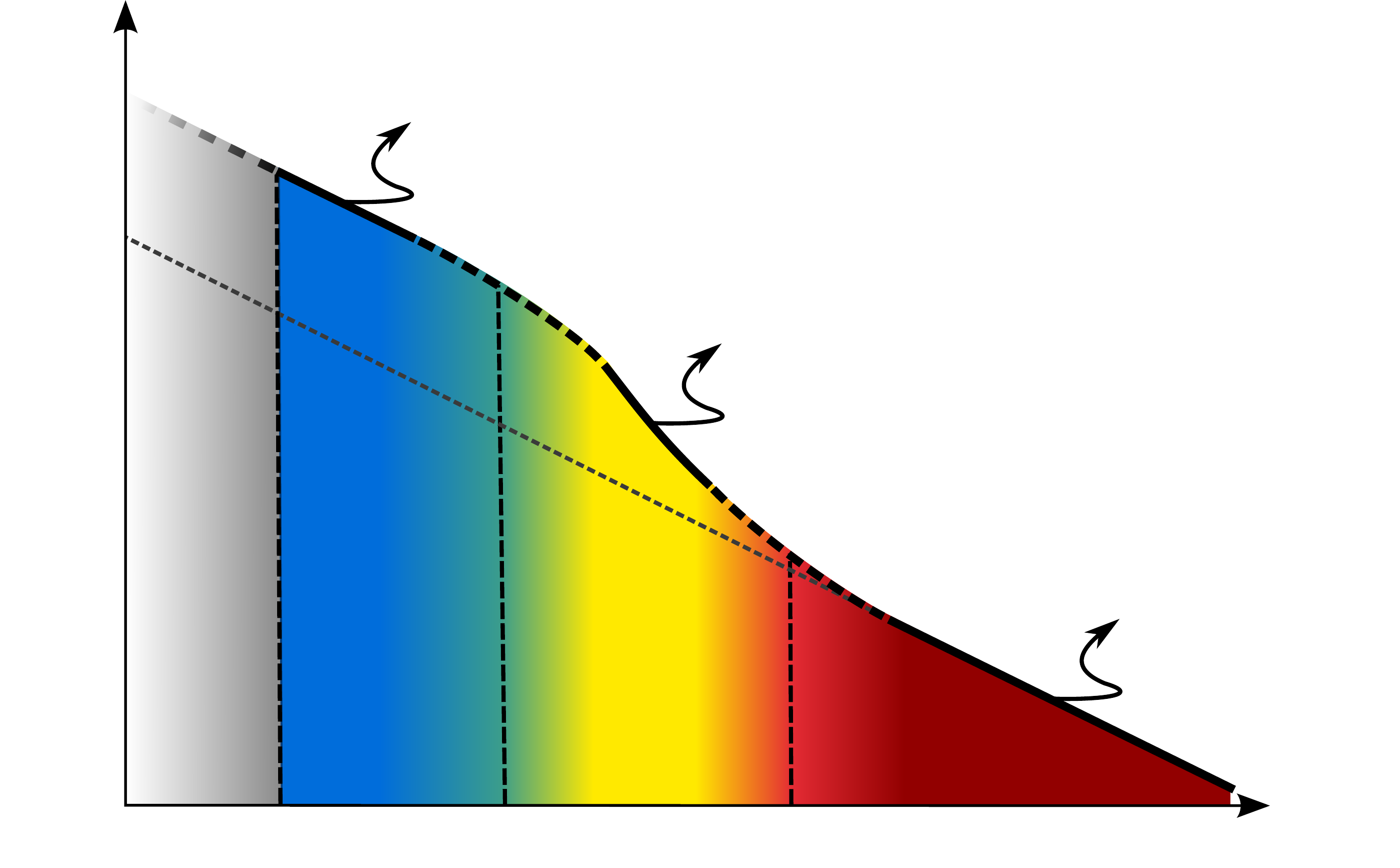}}%
    
   \put(0,0.59){\color[rgb]{0,0,0}\makebox(0,0)[lb]{{{$\ln\! v$}}}}%

   \put(0.54,-0.011){\color[rgb]{0,0,0}\makebox(0,0)[lb]{$H_{r}^{-1/4}$}}%
   
   \put(0.34,-0.01){\color[rgb]{0,0,0}\makebox(0,0)[lb]{$H_{\varepsilon}^{-1/4}$}}%
   
   \put(0.18,-0.02){\color[rgb]{0,0,0}\makebox(0,0)[lb]{$H_{d}^{-1/4}$}}%

   \put(0.85,-0.0125){\color[rgb]{0,0,0}\makebox(0,0)[lb]{$H_{ }^{-1/4}$}}%

        \put(0.30,0.57){\color[rgb]{0,0,0}\makebox(0,0)[lt]{\begin{minipage}{0.66511765\unitlength}\raggedright ${\sim}v_0\varepsilon^{-\nu(\zeta + 1)}$\end{minipage}}}%

               \put(0.52,0.41){\color[rgb]{0,0,0}\makebox(0,0)[lt]{\begin{minipage}{0.85243391\unitlength}\raggedright ${\sim}v_0\delta h^{-\nu(\zeta+1)}$\end{minipage}}}%
        
             \put(0.805,0.19){\color[rgb]{0,0,0}\makebox(0,0)[lt]{\begin{minipage}{0.40146739\unitlength}\raggedright ${\sim}v_0$\end{minipage}}}%
        
\end{picture}
\endgroup
\caption{Three different velocity-field characteristics identified below $H_d$: the classical creep regime for fields $0 < H < H_r$, where the velocity goes as $v_0 e^{-T_d/T (H/H_d)^{-\mu}}$; the field-dependent correction to the velocity prefactor coming from the relaxation contribution gives the excess velocity for fields $H_r < H < H_{\varepsilon}$; and the saturation of the relaxation length at $L_\varepsilon$, for fields $H_{\varepsilon} < H < H_d$, arrests the divergence of the velocity with an $\varepsilon$-dependent velocity prefactor.}
\label{fig:regimes}
\end{center}
\end{figure}
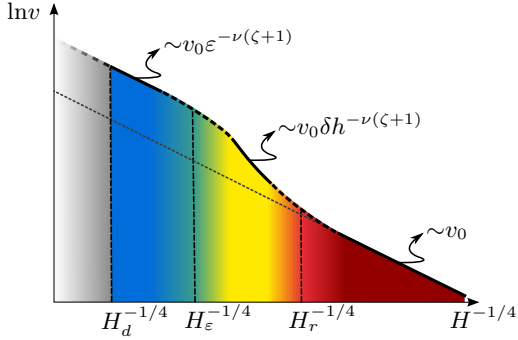

All the information regarding the field-dependence of relevant length scales can be put together by writing the velocity prefactor $u/\tau_0$ in terms of the relaxation contribution and its saturation value, using $(\delta h^m + \varepsilon^m)^{-\nu/m}$ instead of $\delta h^{-\nu}$ in Eq.~\eqref{eq:veldh}, resulting in the corrected- creep law:
\begin{eqnarray}
\label{eq:corrected-creep law}
 v &=& v_0 \left\{ 1 + p \left[ \left( \delta h^m + \varepsilon^m \right)^{-\nu/m}-1\right]^{\zeta+1} \right\} \\ \nonumber 
 & & \times e^{-\frac{T_d}{T}\left[ \left(1-\delta h\right)^{-\mu}-1\right]}.
\end{eqnarray}
Summarizing, this phenomenological corrected creep model includes three different velocity-field characteristics below the depinning field (see Fig. \ref{fig:regimes}).

Increasing the field from zero, when $H < H_r$, the usual creep-law of the form
 \begin{equation}
 v = v_0  e^{-\frac{T_d}{T} \left[ \left( \frac{H}{H_d} \right)^{-\mu} -1 \right]}
\end{equation}
is recovered. Then, in the range $H_r < H < H_\varepsilon$, the relaxation length scales modify the velocity prefactor and the excess velocity develops:
\begin{equation}
 v = v_0 p \left( \frac{H_d-H}{H_d} \right)^{-\nu (\zeta+1)} e^{-\frac{T_d}{T} \left[ \left( \frac{H}{H_d} \right)^{-\mu} -1 \right]}.
\end{equation}
Finally, when the relaxation length reaches its saturation value for $H_\varepsilon < H < H_d$, a modified creep-law with an $\varepsilon$-dependent prefactor is recovered,
\begin{equation}
 v = v_0 p \varepsilon^{-\nu (\zeta+1)} e^{-\frac{T_d}{T} \left[ \left( \frac{H}{H_d} \right)^{-\mu} -1 \right]}.
\end{equation}
These three velocity-field regimes of creep motion are indicated in Fig.\ref{fig:regimes}, together with the characteristic fields $H_r$ and $H_\varepsilon$.

\subsection{Fitting velocity curves}
\label{sec:fits}

\begin{figure}[]
\begin{center}
\includegraphics[width=0.9\columnwidth]{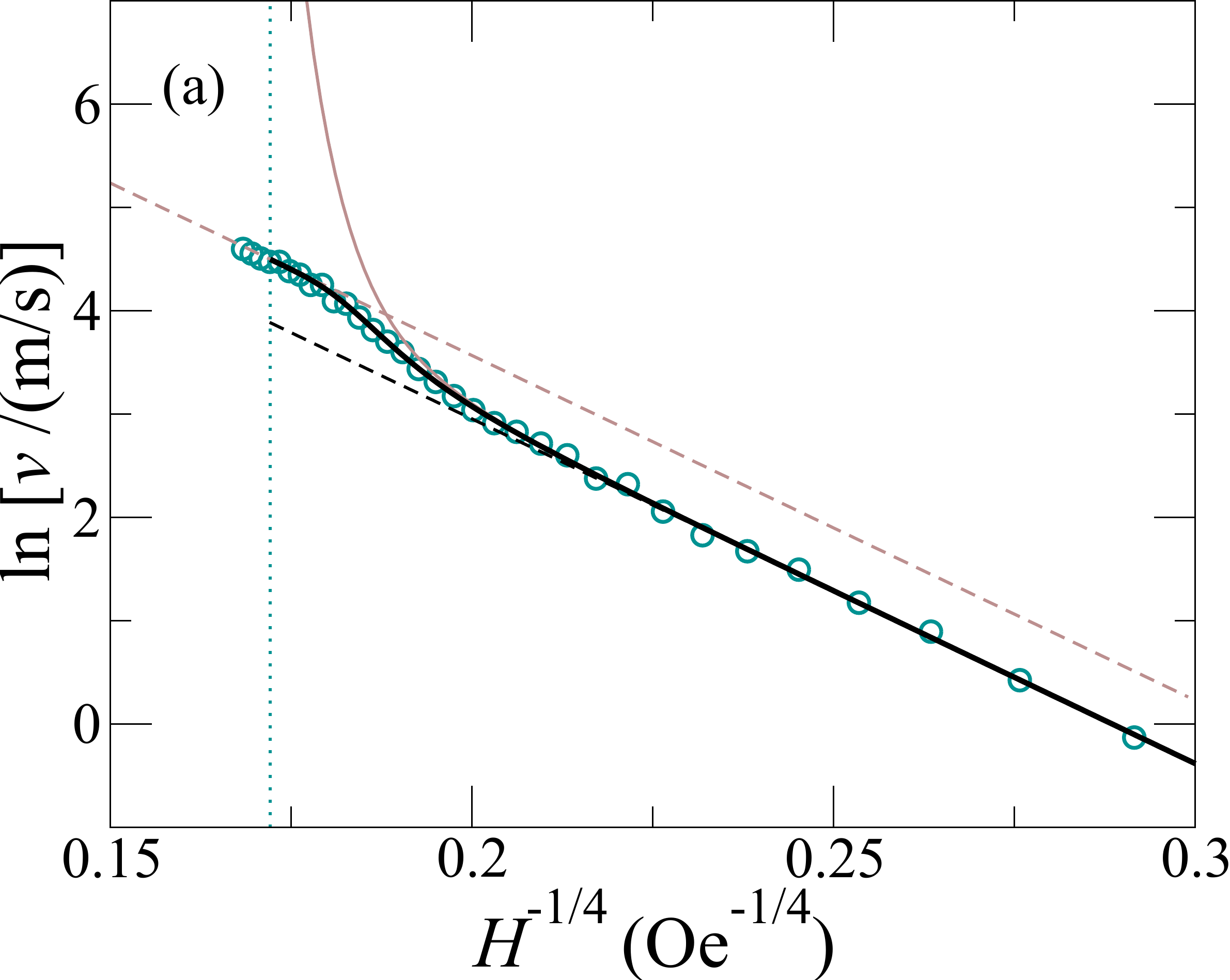}
\includegraphics[width=0.9\columnwidth]{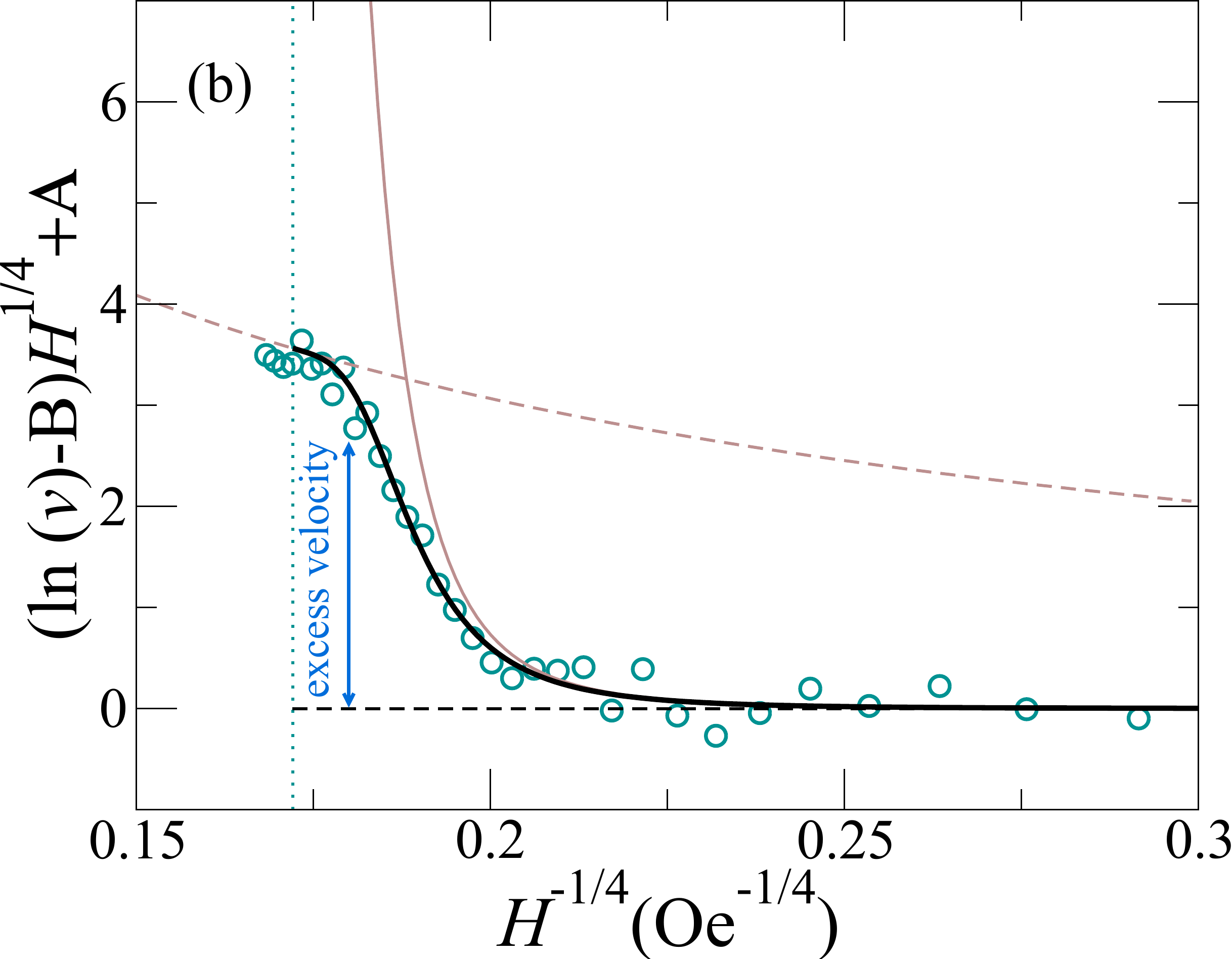}
\caption{(a) Velocity-field characteristics in a creep plot, $\ln v$ against $H^{-1/4}$, for Pt/Co/Pt at room temperature~\cite{Gorchon2014}. The usual creep law (black dashed lines) fits the low magnetic field data. The continuous black line corresponds to a fit using the corrected creep law, which permits to extract the parameters shown in Table~\ref{tab:fit-Gorchon}. The corrected creep law was fit using fields $H<H_d$ (vertical dotted line). The thin continuous line shows the divergence of the velocity due to the divergence of $S_{rel}$, plotted using the same parameters of Table~\ref{tab:fit-Gorchon} but with $\varepsilon=0$. Finally, the upper straight dashed line shows the effect of the saturation which gives an extra contribution to the prefactor $v(H_d)$ very close to $H_d$, Eq.~\eqref{eq:vhd-corr}.
(b) Same data as in (a) but with a flattened low field creep regime (see Eq.\eqref{eq:lnv-creep}), which emphasizes the excess velocity.}
\label{fig:fit-Gorchon}
\end{center}
\end{figure}

In the following we shall use the proposed model to fit velocity-field curves for [Co/Ni] multilayers and extract relevant parameters. For the purpose of the fit, the modified creep-law, Eq.~\eqref{eq:corrected-creep law} can also be written as
\begin{eqnarray}
\label{eq:lnv-corrected}
 \ln v &=& \ln v_0 + \frac{T_d}{T} - \frac{T_d}{T} \left(\frac{H}{H_d} \right)^{-\mu} \nonumber \\
 & &+ \ln \left( 1 + p \left\{ \left[ \left( \frac{H_d-H}{H_d} \right)^m + \varepsilon^m \right]^{-\nu/m} -1 \right\}^{\zeta+1} \right), \nonumber \\
\end{eqnarray}
highlighting how the excess velocity correction to the usual creep comes from the field-dependence of the last term.
Inspecting the corrected creep model Eq.~\eqref{eq:lnv-corrected} one can see that the usual creep model, first line in Eq.~\eqref{eq:lnv-corrected}, is also recovered.

In order to reduce the number of fitting parameters we must first fix the values of the critical exponents, by consistently assuming an equilibrium and a depinning universality class. For domain wall motion in ultrathin ferromagnetic films it is usually found that $\mu=1/4$. This value corresponds to the equilibrium universality class of a one dimensional interface with short-ranged elasticity coupled to a short-range correlated random-bond disorder, as $\mu=(2\zeta_{eq}-1)/(2-\zeta_{eq})$, with $\zeta_{eq}=2/3$. For the depinning universality class, in agreement with previous experimental results~\cite{Gorchon2014, Jeudy2016, DiazPardo2017}, we take the quenched Edwards-Wilkinson universality class and thus set $\nu=1/(2-\zeta)=1.33$ and $\zeta=1.25$~\cite{Ferrero_PRE_2013_depinning}.

As a first step, we validate our model using room temperature velocity-field data obtained for Pt/Co/Pt in Ref.~\cite{Gorchon2014}. This results show the same upward deviation from the full creep law when increasing the field and approaching the depinning field from below. In addition, in the same experiment, the depinning field was reached, giving access to depinning parameters [$H_d$, $T_d$ and $v(H_d)$]. Experimental results are fit using the corrected creep law Eq.~\eqref{eq:corrected-creep law}.
In the low field regime, $H < H_r$, Eq.~\eqref{eq:lnv-corrected} reduces to
\begin{equation}
\label{eq:lnv-creep}
 \ln v = \ln v_0 + \frac{T_d}{T} - \frac{T_d}{T} \left(\frac{H}{H_d} \right)^{-\mu} = B - A H^{-\mu}.
\end{equation}
Therefore we plot the data as a creep plot, $\ln v$ against $H^{-1/4}$, in Fig.~\ref{fig:fit-Gorchon} (a). As can be observed by the agreement between the data and the continuous black line, the proposed model gives a fairly good description of the phenomenon. Figure~\ref{fig:fit-Gorchon} (b) shows the same information with the low field creep law linearized as $(\ln v - B)H^\mu+A$ against $H^{-\mu}$, in order to highlight the excess velocity. Notice that the correction to the velocity prefactor, when the saturation of value $S_\varepsilon$ is reached, gives the velocity exactly at the depinning field,
\begin{equation}
\label{eq:vhd-corr}
 v(H_d)=v_0 \left[1+p \left(\varepsilon^{-\nu }-1 \right)^{\zeta+1} \right],
\end{equation}
corresponding to the prefactor for the gray dashed line in Fig.~\ref{fig:fit-Gorchon}.

In Fig.~\ref{fig:fit-Gorchon}, the thin continuous line shows the divergence of the velocity due to the divergence of $S_{rel}$, plotted using the same parameters of Table~\ref{tab:fit-Gorchon} but with $\varepsilon=0$. Finally, the upper straight dashed line shows the effect of the saturation which gives an extra contribution to the prefactor $v(H_d)$ very close to $H_d$, Eq.~\eqref{eq:vhd-corr}.
In Table~\ref{tab:fit-Gorchon}, fitting parameters using the corrected creep law model for $H<H_d$ are compared with the values obtained using the usual creep law for small magnetic field values in Ref.~\cite{Gorchon2014} (see also Ref.~\cite{Jeudy2017_parameters} for a comparison with other materials).

\begin{table}[]
\centering
\begin{tabular}{lcccccc}
\hline
\hline
             & Usual creep & Corrected creep\\
             &Ref.~\cite{Gorchon2014,Jeudy2017_parameters}& \\
\hline
$H_{d}$[Oe]  & 910$\pm$40  & 1140$\pm$60\\
$T_{d}$[K]   & 1900$\pm$100& 1720$\pm$50\\
$v_0$[m/s]   & 59$\pm$2    & 48$\pm$18\\
$\varepsilon$&             & 0.30$\pm$0.04\\
$p$          &             & 0.04$\pm$0.02\\
$v(H_d)$[m/s]& 59$\pm$2    & 90$\pm$35\\
\hline \hline
\end{tabular}
\caption{Fitting parameters for Pt/Co/Pt velocity-field data at room temperature (data from Ref.~\cite{Gorchon2014}). Comparison between the fitting parameters obtained using the usual creep (values extracted from Ref.~\cite{Jeudy2017_parameters}) and using the corrected creep law for $H<H_d$ proposed in this work, Eq.~\eqref{eq:lnv-corrected}.}
\label{tab:fit-Gorchon}
\end{table}


Figure~\ref{fig:M1M2M3}(b) shows experimental velocity-field curves for all three Co/Ni samples in a creep plot and fitting curves using both the usual creep law and the corrected creep law, Eqs.~\eqref{eq:lnv-creep} and \eqref{eq:lnv-corrected}, respectively. As can be observed, the corrected creep expression, Eq.~\eqref{eq:lnv-corrected}, gives a very good agreement on the whole studied experimental range. The obtained results for the fitting parameters are presented in Table~\ref{tab:parametros}.
It is worthwhile to emphasize here that solely using the usual creep law is not enough to obtain the three depinning parameters and information from the depinning regime is needed~\cite{Jeudy2016, Jeudy2017_parameters}. Remarkably, whenever the upward velocity deviation is present, the proposed corrected creep model permits to extract the full set of parameters using data well below the depinning field, as can be observed for NiCoAl and CoNiAl in Fig.~\ref{fig:M1M2M3}(b).

Overall, data presented in Fig.~\ref{fig:M1M2M3} and results in Table~\ref{tab:parametros} show that significant differences are observed among the studied Co/Ni samples. These differences are probably related to the presence of different interfaces between the capping/buffer layers and the magnetic multilayer. Furthermore, one can notice in Table~\ref{tab:parametros} that while $p$ changes, the value of $\varepsilon$ does not change within error bars, suggesting a complicated scenario which should be unveiled. We expect the temperature dependence of these parameters would reveal more details about their relationship with materials intrinsic parameters.

\begin{table}[]
\centering
\begin{tabular}{lccc}
\hline
\hline
	     & NiCoAl & CoNiTa & CoNiAl\\
\hline
$H_{dep}$[Oe]& 110$\pm$35   & 210$\pm$30 & 820$\pm$200\\
$T_{dep}$[K] & 3900$\pm$330 &3100$\pm$130 & 2300$\pm$150\\
$v_0$[m/s]   & 1.7$\pm$0.5 &0.25$\pm$0.05 & 0.09$\pm$0.01\\
$\varepsilon$& 0.6$\pm$0.2 &0.47$\pm$0.09 & 0.4$\pm$0.1\\
$p$          & 2.8$\pm$0.8 &1.5$\pm$0.5 & 20$\pm$7\\
$v(H_d)$[m/s]& 6$\pm$1 &1.4$\pm$0.5 & 12.8$\pm$0.7\\
\hline \hline
\end{tabular}
\caption{Fitting parameters of the three studied [Co/Ni] multilayers, obtained from fitting the measured velocities with the corrected full creep law, Eq.~\eqref{eq:lnv-corrected}. For further reference, we also included $v(H_d)$ from Eq.~\eqref{eq:vhd-corr}.}
\label{tab:parametros}
\end{table}

\section{Discussion and conclusion}
\label{sec:discussion}

We have found that the excess velocity effect can be explained using a simple phenomenological model which incorporates the relaxation contribution to the creep events near the depinning field. We present here a discussion regarding various specific aspects of the model.

The model predicts that the excess velocity is controlled by a mix of equilibrium and depinning exponents. In section \ref{sec:fits} we have fixed the equilibrium and depinning universality class to the one of the quenched Edwards-Wilkinson model. For depinning, this model predicts $\zeta \approx 1.25$, and $\nu=1/(2-\zeta)\approx 4/3$ using the statistical-tilt symmetry of the model \cite{fisher1991,chauve_creep_long}. The roughness exponent $\zeta=1.25 > 1$ implies however that the local elongation must diverge with the domain wall segment size. Such divergence is incompatible with the harmonic elastic approximation of the model at large length-scales \cite{thermal_rounding_fitexp}.
This may imply that either the interface breaks, giving place to plastic flow at large length-scales, or non-linear corrections to the elasticity constraining the local elongation become important. The latter is the case studied in Ref.~\cite{rosso_hartmann} for a non-harmonic quartic order constraint to the local elongation. It was shown that in this case the depinning universality class corresponds to the quenched Kardar-Parisi-Zhang depinning universality class. Interestingly, the equilibrium universality class of this model is the same as for the quenched Edwards-Wilkinson model with $\zeta_{eq}=2/3$ and in particular $\mu=1/4$. Therefore one can not distinguish between the two cases just by measuring the mean velocity at low fields. One may thus ask whether the excess velocity regime may reveal the depinning universality class. We have thus tested our model using the known quenched Kardar-Parisi-Zhang depinning exponents, $\zeta \approx 0.63$ and $\nu \approx 1.75$. The result shows that the model still fits satisfactorily the data for the Pt/Co/Pt sample (the other samples give similar results), and the fitting parameters do not change appreciably.
Therefore, our experiment and velocity model are not able to resolve the depinning universality class of the magnetic domain wall system, at least between the quenched Edwards-Wilkinson and quenched Kardar-Parisi-Zhang models.

When the depinning field is approached from above, $H > H_d$, the depinning correlation length $L_{av}$ is the characteristic crossover length scale beyond which the geometry of the interface is well described by the fast-flow exponent $\zeta=1/2$ and diverges as $L_{av} \sim (H-H_d)^{-\nu}$, when $T=0$. The time scale it takes to such depinning correlation length to develop diverges as $t_{av} \sim L_{av}^z$, with $z$ the depinning dynamical exponent~\cite{kolton_short_time_exponents, Ferrero_PRE_2013_depinning}. In principle one can analyze the depinning in analogy with standard equilibrium critical phenomenon~\cite{fisher_depinning_meanfield}, suggesting the existence of a diverging length scale below $H_d$ and its corresponding time scale. As shown in Refs.~\cite{kolton_depinning_zerot2, kolton_dep_zeroT_long} this analogy breaks down and although $L_{rel}$ diverges when $H_d$ is approached from below, Eq.~\eqref{eq:Lreldiv}, it is associated with transient dynamics. Furthermore, the corresponding time scale $t_{rel} \sim L_{rel}^z$ should also diverge at $H_d$. This time scale is, at very small fields, smaller than the time it takes to overcome the typical energy barrier, $t_{rel} < \tau$. If one considers the divergence of $t_{rel}$, this would lead to a depinning-like power-law behavior for the velocity but not to a speeding up of the velocity over the usual creep law. Therefore, when developing the corrected creep model in Sec.~\ref{sec:corrected-model} we had assumed that for $H < H_\varepsilon$ the time scale $t_{rel} \ll \tau$, thus not giving an important time correction. Furthermore, when $L_{rel}$ saturates for $H > H_\varepsilon$, so does the value of $t_{rel}$, not affecting the velocity prefactor.

From the proposed model and the arguments developed in Sec.~\ref{sec:corrected-model}, it is clear that in order to avoid the divergence of the velocity a saturation of $L_{rel}$ is necessary when approaching $H_d$ from below. The possible origin of the saturation length $L_\varepsilon$ needs further discussion, with one interesting candidate being the depinning correlation length $L_{av}$. 
It would be very interesting to test experimentally and numerically the field and temperature implications of considering $L_\varepsilon = L_{av}$, as it might provide a possible link between thermal rounding, the density of concurrent creep events, and the effective barrier distribution. For instance, the characteristic length $L_{av}$ is field dependent and at $T=0$ it diverges at $H_d$ while
at any finite temperature is finite at $H_d$ and tends to diverge at $H=0$ \cite{chauve_creep_long,kolton_universal_aging_at_depinning}. The larger is $T$ the smaller is $L_{av}(H)$. Associating $L_{av}$ with $L_\varepsilon$ means that the saturation regime would be reached at lower fields. On the other hand, the onset of the excess velocity regime $H_r$ is also expected to decrease. Depending on the precise temperature dependence of both, the excess velocity window may shift, expand or shrink, and eventually disappear. Measuring the temperature dependence of the velocity field characteristics below $H_d$ would allow to test this picture experimentally. 

In summary, we measured domain wall velocities driven by magnetic fields in [Co/Ni] based multilayers using PMOKE microscopy.
In the three samples analyzed the low field creep regime is observed in agreement with the classical creep law, $\ln v \sim H^{-1/4}$.  When increasing the field an unexpected speeding up of the velocity, an upward deviation from the creep law, is observed. This upward trend is also compatible with previous measurements, for example in Refs.~\cite{Gorchon2014, LeGall2015}. Using a phenomenological approach based on scaling arguments we were able to obtain a large field correction to the low field creep law which explains the observed behavior. The correction includes the field dependence of the transverse displacement associated to relaxation events and a saturation cut-off very close to $H_d$, thus arresting the divergence of the velocity.
The proposed model permits to predict three velocity-field characteristics below the depinning field, from $H=0$ up to $H_d$, and to obtain the relevant depinning parameters ($H_d$, $T_d$, $v(H_d)$) by fitting the experimental data. Despite the fact that only data for $H<H_d$ is used in the fitting procedure, our model gives very good agreements with previously obtained depinning parameters.
We hope to motivate further experimental and theoretical research. In particular, a detailed experimental analysis of the field and temperature dependence of the velocity below the depinning threshold, and the theoretical study of the density of activated events in the creep regime would be welcome.

\acknowledgements

A. B. K. acknowledges discussions with E. A. Jagla, A. Rosso and L. Foini. This work was partly supported by the Argentinian projects PIP11220120100250CO (CONICET), and UNCuyo Grants No. 06/C490 and 06/C017. This work was also partly supported by the Labex NanoSaclay ANR-10-LABX-0035. S. B., N. C., J. C. and V. J. acknowledge support by the France-Argentina project ECOS-Sud No. A12E03. R. D. P. thanks the Mexican council CONACyT for the PhD fellowship No.: 449563.

\bibliography{tfinita5,paredom,dmi}

\end{document}